\documentclass[letterpaper,twocolumn,10pt]{article}
\usepackage{usenix2019_v3}
\usepackage{xspace}
\usepackage{graphicx}
\usepackage{float}
\usepackage{subcaption}
\usepackage{amsmath}

\usepackage{amssymb}
\usepackage{array}
\usepackage{bbding}
\usepackage{makecell}
\usepackage{booktabs}
\usepackage{multirow}
\usepackage{colortbl}
\usepackage{enumitem}
\usepackage{siunitx}
\usepackage{listings}
\usepackage{xcolor}

\usepackage{amsfonts}
\usepackage{algorithmic}
\usepackage{textcomp}
\usepackage{tagging}
\usepackage[most]{tcolorbox}
\usepackage{tikz}
\usepackage{tikz}
\usepackage{amsmath}
\usepackage{booktabs}

\usepackage{filecontents}

\newcommand{\sys}{\textsc{Coda}\xspace}

\newcommand{\revadd}[1]{\textcolor{black}{#1}}

\newtcolorbox{findingbox}{
  breakable,
  colback=black!3,
  colframe=black!60,
  boxrule=0.45pt,
  arc=1pt,
  boxsep=0pt,
  left=4pt,
  right=4pt,
  top=3pt,
  bottom=3pt,
  before skip=4pt,
  after skip=5pt,
  fontupper=\normalsize
}

\begin{document}

\date{}

\title{\sys: Exploiting Admission Flexibility for Coding-Agent Serving}

\author{
{\rm Youhe Jiang}\\
University of Cambridge
\and
{\rm Fangcheng Fu}\\
SJTU
\and
{\rm Binhang Yuan}\\
HKUST
\and
{\rm Krishna Malladi}\\
Meta
\and
{\rm Ehsan K. Ardestani}\\
Meta
\and
{\rm Zhan Shu}\\
Meta
\and
{\rm Adnan Aziz}\\
Meta
\and
{\rm Yi Xu}\\
Meta
} 

\maketitle

\begin{abstract}
Coding agents powered by large language models (LLMs) repeatedly alternate between
model inference and tool calls, creating long-lived sessions with reusable
key-value (KV) states and asynchronous request resumptions. \revadd{Logical readiness},
however, does not ensure efficient admission in a shared serving system.
Through direct trace analysis and trace-driven replay, we identify two
mismatches: reusable KV states reside across storage tiers and incur unequal
\revadd{computational costs for state preparation}, while requests with heterogeneous context lengths can
decode inefficiently together. Our central insight is that exploiting
admission flexibility can improve serving performance while preserving
request progress.

We present \sys, a coding-agent serving system that realizes this insight
through a readiness-informed admission layer incorporating two mechanisms.
\textit{Tiered-Aging state admission} exploits \revadd{bounded} flexibility in \revadd{admission order} to improve admission efficiency and preserve request
progress. \textit{Compatibility-Aware execution admission} exploits
flexibility in \revadd{attention grouping within each model iteration} to reduce mixed-context interference
and improve shared decoding efficiency.
For multi-worker configurations, \sys introduces a separate routing layer
that considers KV-state residency, worker load, and request-to-worker
\revadd{context-length compatibility} to guide efficient request placement.
We evaluate \sys
across different models and workloads against state-of-the-art coding-agent
serving systems and vLLM. Across the single-worker and multi-worker
experiments, \sys improves output-token and SLO-compliant throughput by
20.3\% and 70.5\% on average, with peak gains of 29.3\% and 140.2\%.
\end{abstract}

\section{Introduction}
\label{sec:introduction}

\begin{figure}
    \centering
    \includegraphics[width=0.95\linewidth]{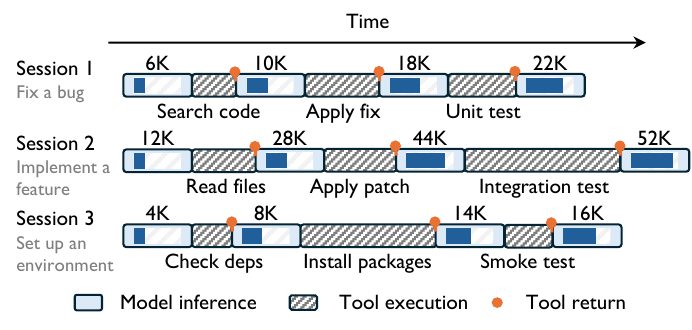}
    \caption{\textbf{Illustrative multi-session coding-agent workflows.}
Independent sessions alternate between model inference and tool execution.
Tool calls return asynchronously, and their outputs extend subsequent
model inputs, causing session contexts to grow at different rates.}
    \label{fig:aria-intro-pic}
\end{figure}

LLM coding agents extend software automation from code generation to
multi-step development workflows, including repository-level repair, code
editing, environment configuration, and test
execution~\cite{wang2025openhands,bouzenia2025you,yang2024swe,zhu2026tracelab}.
Systems such as OpenAI Codex~\cite{openai_codex}, Claude
Code~\cite{anthropic_claude_code}, and Cursor~\cite{cursor_agent} support
these workflows through integrated model inference and tool execution.
As \autoref{fig:aria-intro-pic} illustrates, concurrent sessions repeatedly
pause for tools, resume asynchronously, and accumulate context across model
rounds~\cite{tiwari2026cachewise,xia2026idleness}. A request becomes
\textit{logically ready} when user input arrives or a tool return enables its
session to continue. Logical readiness determines whether a request may
proceed, but it does not determine whether that request should \revadd{be prepared next} or
how efficiently it will share GPU execution with other ready requests.
Admitting a request commits GPU resources and changes how other requests share
those resources. Two characteristics of coding-agent workloads make this
distinction particularly important.

\textbf{First, readiness does not imply equal computational cost for preparation.}
Long-lived coding-agent sessions accumulate reusable interaction histories
across successive model and tool invocations. Preserving the KV
cache avoids repeated prefix computation, but the aggregate KV state of
concurrent sessions can exceed available GPU high-bandwidth memory (HBM)
capacity, motivating
storage across HBM, CPU DRAM, and lower storage
tiers~\cite{lmcache-gds,qin2026mooncake}. During tool-induced idle periods,
memory pressure may evict a session's KV state from HBM before its next
request becomes ready~\cite{xia2026idleness,tiwari2026cachewise,kang2026thunderagent}.
Consequently, equally ready requests can require substantially different
preparation: one may reuse HBM-resident state, whereas another requires
restoration, storage access, or prefix recomputation. In our trace-driven
replay (\S\ref{sec:trace-characterization}),
24.45\% of resumed requests are not HBM-ready but account for 42.45\% of
total \revadd{startup computational cost}. Admission based on arrival order or reusable-prefix
availability alone can therefore allow computationally expensive state preparation to delay
several computationally inexpensive session continuations.

\textbf{Second, readiness does not imply efficient shared execution.}
Long-lived coding-agent sessions accumulate context at different rates, while
asynchronous tool returns bring sessions at different stages back into
execution~\cite{zhu2026tracelab,tiwari2026cachewise}. Consequently,
short-context requests frequently execute alongside requests with
substantially larger KV states. Although each request is ready to proceed,
heterogeneous context lengths can reduce attention-kernel efficiency during
shared decoding~\cite{bai2026alignedserve}. In our trace-driven replay
(\S\ref{sec:trace-characterization}), 82.84\% of multi-request decode
iterations have a largest-to-smallest context-length ratio of at least
4$\times$. \revadd{For pure-decode iterations, at} fixed request count and total context-token volume, the observed
context composition incurs a 25.95\% higher median computational cost per iteration than a
near-uniform composition. Admission based only on individual readiness can
therefore create a heterogeneous active population; treating that population
as a single attention group leaves \revadd{context-length compatibility} unaddressed.

Exploiting these opportunities is challenging. Prioritizing requests with low
computational costs of state preparation can reduce blocking, but may repeatedly bypass sessions
whose KV states reside in lower tiers. Conversely, strict ready-time order
prevents reordering-based bypass but sacrifices this efficiency opportunity.
Similarly, organizing heterogeneous contexts into separate attention groups
can improve attention efficiency, but additional launches and smaller groups
may outweigh the benefit; this tradeoff must be assessed online as the active
population changes. In multi-worker configurations, context-compatible placement
can also conflict with KV affinity and load balance.

\begin{findingbox}
\textit{Logical readiness should determine eligibility, but not admission
order or execution grouping. Our approach exploits bounded flexibility between
readiness and execution to reorder state preparation and reorganize attention
groups.}
\end{findingbox}

In this paper, admission is twofold: state admission determines which logically
ready request is prepared next, while execution admission determines how the
current participating requests are organized for execution. These two decisions
expose flexibility in admission order and execution grouping:

\textbf{Flexibility in admission order.} Admission can account for
differences in the computational costs of KV-state preparation when selecting which logically ready
request is prepared next. Bounded reordering can improve admission efficiency
while guarding against indefinite bypass.

\textbf{Flexibility in execution grouping.} Admission can account for
context-length compatibility when organizing the participating requests for
attention execution. Compatibility-aware grouping can improve shared decoding
efficiency when its computational savings exceed the computational cost of additional execution groups.

Existing systems improve coding-agent serving through cache management,
session placement, and resource coordination. CacheWise combines
prefix-aware scheduling based on additional KV-block demand with predictive
KV-cache eviction informed by tool calls~\cite{tiwari2026cachewise}. SMetric
uses session reuse and worker load to guide request
routing~\cite{wang2026smetric}, while MORI manages GPU and CPU residency
through relative program idleness and tier
admission~\cite{xia2026idleness}. AlignedServe instead forms batches from
waiting requests with similar KV-cache
lengths~\cite{bai2026alignedserve}. Existing systems therefore address individual aspects of residency,
placement, or context-aware batching, but do not jointly provide bounded
preparation-aware ordering and within-iteration attention grouping for
asynchronously resuming coding-agent sessions.

We present \sys, a coding-agent serving system that decouples logical
readiness from physical GPU execution through a readiness-informed admission
layer. \textit{Tiered-Aging state admission} realizes bounded admission-order
flexibility by combining the computational cost of KV-state preparation, waiting age, and a progress
guard. \textit{Compatibility-Aware execution admission} realizes
execution-grouping flexibility by applying context-compatible attention groups
only when conservative predicted computational savings exceed their additional computational overhead. For
multi-worker configurations, a separate routing layer considers context-length
compatibility only within KV-affinity and load safeguards. Our contributions
are summarized as follows:

\noindent\textbf{\underline{Contribution 1:}} We characterize coding-agent workloads
through direct trace analysis and replay
(\S\ref{sec:trace-characterization}). We examine how changing KV-state
residency creates unequal \revadd{computational costs for state preparation}, and how uneven context
growth and tool returns create heterogeneous decode contexts.
These findings show why logical readiness alone is insufficient and motivate
flexible admission order and \revadd{execution grouping}.

\noindent\textbf{\underline{Contribution 2:}} We design and implement \sys with two
admission mechanisms (\S\ref{sec:system-design-1}): Tiered-Aging state admission
(\S\ref{subsec:tiered-aging-state-admission}) and Compatibility-Aware execution
admission (\S\ref{subsec:compat-aware-execution-admission}). A complementary
routing layer supports multi-worker configurations (\S\ref{sec:system-design-2}). We
integrate these mechanisms as an orchestration layer connected to the inference
backend (\S\ref{sec:implementation}).

\noindent\textbf{\underline{Contribution 3:}} We evaluate \sys across models and workloads
in single-worker and multi-worker configurations (\S\ref{sec:evaluation}).
Compared with state-of-the-art coding-agent serving systems and a vLLM
baseline, \sys improves single-worker output-token throughput by 17.5\%
on average and up to 23.9\%, and SLO-compliant throughput by 96.1\% on
average and up to 140.2\%. In the multi-worker experiments, output-token
throughput improves by 23.1\% on average and up to 29.3\%, while
SLO-compliant throughput improves by 45.0\% on average and up to 59.0\%.
Component ablations and runtime analysis further identify the contributions
of Tiered-Aging state admission, Compatibility-Aware execution admission,
and routing to serving performance.

\section{Background and Motivation}
\label{sec:background}

\noindent\textbf{Coding-agent serving background.}
Coding-agent workflows interleave model inference with development tools
for repository inspection, code modification, command execution, and
testing~\cite{yang2024swe,wang2025openhands}. A session comprises multiple
dependent model requests, with tool outputs incorporated into the input
of subsequent requests~\cite{yao2022react,tiwari2026cachewise}. Each model request involves prefill computation
for input tokens and iterative decoding for output generation.
Retaining the corresponding key-value (KV) state allows subsequent
requests to reuse the session prefix rather than repeatedly compute
it~\cite{gao2024cost,liu2025lmcache}. Unlike isolated inference requests, coding-agent sessions retain
growing interaction histories and repeatedly pause for tool execution,
producing long-lived KV states and asynchronous
resumptions~\cite{zhu2026tracelab,liu2026agentic,chang2026llm}. Concurrent sessions compete for GPU
memory, motivating KV storage across GPU HBM, CPU DRAM, and lower
storage tiers~\cite{liu2025lmcache,qin2026mooncake}. Meanwhile, serving engines combine requests into batches,
so requests from different session stages may decode
together~\cite{kwon2023efficient,yu2022orca}. Admission therefore affects more than
an individual request: \revadd{admission order
determines which request's state-preparation work occurs next and shapes
which contexts execute concurrently}.

\begin{figure}
    \centering
    \includegraphics[width=0.95\linewidth]{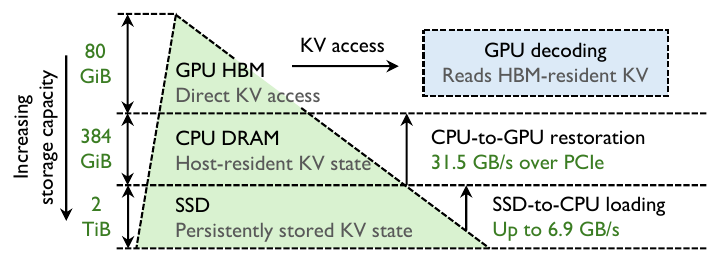}
    \caption{\textbf{Illustrative example of multi-tier KV storage.}
Lower tiers provide larger storage capacity. HBM-resident KV is directly
accessible for decoding, whereas DRAM-resident KV requires GPU restoration
and SSD-resident KV first requires loading into DRAM.}
    \label{fig:memory-tier}
\end{figure}

\noindent\textbf{Multi-tier KV storage.}
GPU serving systems extend KV-cache storage across GPU HBM, CPU DRAM,
and SSDs to balance access efficiency and capacity
across the hierarchy~\cite{sheng2023flexgen,gao2024cost,qin2026mooncake}.
GPU HBM provides high-bandwidth access to KV state during model
execution, but its limited capacity must also accommodate model
weights and runtime buffers~\cite{kwon2023efficient,gao2024cost}.
CPU DRAM provides additional capacity for reusable KV state, while
SSDs offer a larger-capacity storage tier~\cite{liu2025lmcache}.
As illustrated in \autoref{fig:memory-tier}, these tiers differ in the work required to reuse cached state:
HBM-resident KV can be accessed directly, CPU-resident KV requires
transfer to the GPU, and SSD-resident KV additionally requires
storage I/O~\cite{gao2024cost,liu2025lmcache}.
If reusable state is unavailable, the corresponding prefix must be
recomputed. Restoring cached state also does not eliminate prefill
computation for newly appended input tokens. Consequently, retaining
a reusable prefix avoids redundant computation, but its current
storage location still affects the computational cost of resuming a request.

\noindent\textbf{Systems for coding-agent serving.}
Recent systems exploit session structure, tool behavior, and KV reuse
to improve coding-agent serving. CacheWise combines prefix-aware
scheduling with predictive KVCache eviction: it prioritizes requests
requiring fewer additional KV blocks and uses tool information to
predict relative future reuse for eviction
decisions~\cite{tiwari2026cachewise}. SMetric introduces balanced
session-centric scheduling, using actual local-prefix hits to preserve
session locality and load-aware guards to prevent overloaded workers
from retaining requests solely for reuse~\cite{wang2026smetric}.
MORI derives relative program idleness from inference and tool-call
timings, coordinating \revadd{KV-state residency} and program admission across GPU
and CPU tiers~\cite{xia2026idleness}. ThunderAgent represents
multi-turn workflows as LLM programs and coordinates program execution
through memory-pressure-aware pause and restore decisions, together
with tool-environment management~\cite{kang2026thunderagent}. For interactive LLM serving, Bidaw prioritizes SSD KV loading according
to KV size and waiting time to reduce storage-induced request
blocking~\cite{hu2026bidaw}.
\revadd{CacheWise does not combine tier-specific preparation time with a waiting
bound. SMetric does not consider \revadd{context-length compatibility}, and MORI,
ThunderAgent, and Bidaw do not optimize within-iteration attention grouping.
None of these systems jointly addresses admission order and execution
organization.}
\revadd{AlignedServe groups requests with similar prefix lengths by controlling which requests enter the next batch, potentially delaying requests and requiring starvation handling; \sys instead keeps the participating request set unchanged and reorganizes only its attention computation~\cite{bai2026alignedserve}.}
These systems establish the importance of locality, residency, and
program-level resource control in coding-agent and interactive LLM
serving. \sys instead places admission flexibility at the center of its design, which is based on careful coding-agent workload characterizations and jointly considers \revadd{admission order} and execution grouping within a unified admission layer.

\nocite{jiang2023hexgen,jiang2025hexgen,jiang2025thunderserve,jiang2025demystifying,zhang2025lmdeploy}

\noindent\textbf{Motivation.}
In coding-agent serving, ready requests differ in both the work required
to prepare their KV states and their \revadd{context-length compatibility} for shared decoding.
These differences make admission consequential for shared serving
performance. A returning session may require substantial KV restoration
and prefill work, while another can proceed with little preparation.
Meanwhile, asynchronous resumptions bring requests with different context
lengths into shared decoding, where their execution efficiency depends
on how they are organized together.
\revadd{Logical readiness} therefore leaves two important choices: which ready
request should \revadd{have its KV state prepared} next, and how \revadd{the} participating requests
should \revadd{be organized for execution}. Understanding whether these choices offer
meaningful performance opportunities requires careful characterization of the underlying
coding-agent workload behavior. The following section characterizes KV-state residency
\revadd{and startup computational costs (\textbf{Findings~1-3}), within-worker context
composition and decode efficiency (\textbf{Findings~4-7}), and cross-worker
context composition (\textbf{Finding~8}). These findings motivate our
state-admission, execution-admission, and routing mechanisms, respectively.}

\section{Coding-Agent Workload and System Characterization}
\label{sec:trace-characterization}

\noindent \textbf{Trace and replay methodology.} We characterize the complete TraceLab corpus~\cite{zhu2026tracelab}, which
contains 357,161 model rounds, 432,510 tool calls, and 4,281 physical sessions.
We obtain request- and session-level properties directly from the trace. For
runtime state not recorded by TraceLab, including physical KV-tier placement,
concurrent request populations, decode iterations, and worker assignments, we
use policy-neutral trace-driven replays that preserve request tokens, session
order, and observed tool-call intervals. Specifically, we use a common hierarchy replay with a 30-minute warm-up followed by a
two-hour measurement interval. The evaluation configuration consists of four A100-80GB workers, each with 33.91~GiB of
usable HBM KV-cache capacity, 384~GiB of shared DRAM, and a 2-TiB SSD tier.

\noindent\textbf{Reusable KV states are distributed across storage tiers.}
At the end of the measurement interval, HBM holds 21 resident KV-state copies
(129.5~GiB), DRAM holds 57 (383.4~GiB), and SSD holds 316
(2,047.3~GiB). These copy counts overlap because the hierarchy maintains
lower-tier shadows. During the interval, 39,271 requests resume existing
sessions after user input or tool-call returns. Among them, 9,601 (24.45\%)
are not HBM-ready: 6,769 (17.24\%) require CPU-to-GPU restoration from DRAM,
2,763 (7.04\%) require SSD-to-CPU loading followed by GPU restoration, and 69
(0.18\%) require prefix recomputation. The SSD remains near capacity and
performs 1,858 evictions. Lower-tier restoration generates 64,517.2~GiB of
CPU-to-HBM traffic, including 15,705.3~GiB that first crosses the SSD-to-CPU
link.

\begin{findingbox}
\textbf{\underline{Finding 1.}} Reusable coding-agent KV state spans the storage hierarchy.
A reusable prefix may reside below HBM and require substantial state movement
before execution.
\end{findingbox}

\noindent\textbf{KV-state residency changes continuously.}
Asynchronous tool-call returns repeatedly reactivate different coding-agent
sessions, changing which reusable KV states reside in each storage tier.
Between adjacent 10-minute boundaries, 80.00-100.00\% of HBM-resident session
identities, 79.59-91.67\% of DRAM-resident identities, and 34.95-53.24\% of
SSD-resident identities are replaced. Thus, \revadd{KV-state residency} is a transient
property rather than a fixed session attribute.

\begin{findingbox}
\textbf{\underline{Finding 2.}} Reusable \revadd{KV-state residency changes continuously} as
asynchronous tool-call returns reshape storage-tier residency.
\end{findingbox}

\begin{figure}[t]
    \centering
    \includegraphics[width=\linewidth]{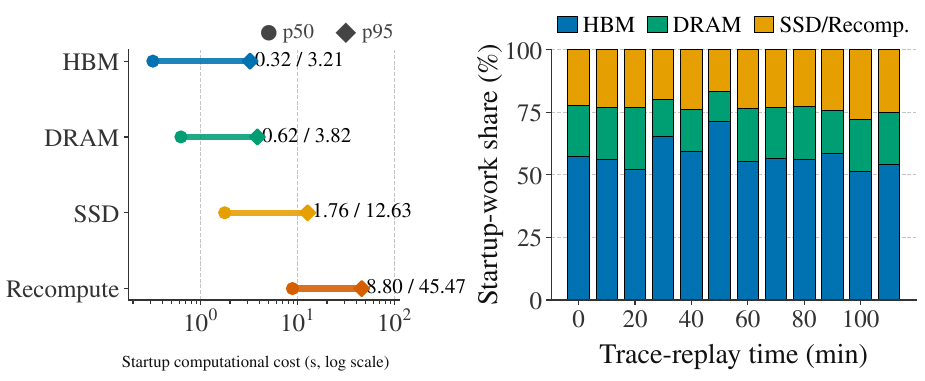}
    \caption{\textbf{\underline{Left:}} p50 (circle) and p95 (diamond) startup computational costs, including KV-state
    movement and prefill, for resumed requests grouped by their required
    KV-state preparation path. \textbf{\underline{Right:}} each stacked bar shows the share of total
    startup work attributable to HBM-ready, DRAM-restoration, and
    SSD-loading/recomputation paths for each successive 10-minute trace-replay
    interval.}
    \label{fig:char-startup-cost}
\end{figure}

\noindent\textbf{Request startup computational costs vary across KV-state preparation paths.}
For each request in the trace replay, \revadd{startup computational cost} captures the KV-state
movement and prefill work required before decoding can begin. The p50/p95
startup computational costs are 0.320/3.214~s for HBM-ready requests, 0.623/3.820~s for
requests requiring DRAM restoration, 1.760/12.633~s for requests requiring SSD
loading, and 8.797/45.466~s for requests requiring prefix recomputation.
Although only 7.21\% of resumed requests require SSD loading or recomputation,
these paths contribute 23.03\% of total startup computational cost. DRAM restoration is
required by 17.24\% of resumed requests and contributes another 19.42\%.
Across successive 10-minute trace-replay windows, requests collectively incur
3,238-4,266~s of startup computational cost. The HBM-ready, DRAM-restoration, and
SSD-loading/recomputation paths contribute 51.22-71.40\%, 11.79-24.78\%,
and 16.81-27.78\% of this computational cost, respectively.
\mbox{\autoref{fig:char-startup-cost}} summarizes the distribution and temporal
composition of request startup computational cost.

\begin{findingbox}
\textbf{\underline{Finding 3.}} Coding-agent request startup computational costs vary substantially
across KV-state preparation paths, and each path's contribution to the
system's startup computational cost changes over time.
\end{findingbox}

\begin{figure}[t]
    \centering
    \includegraphics[width=\linewidth]{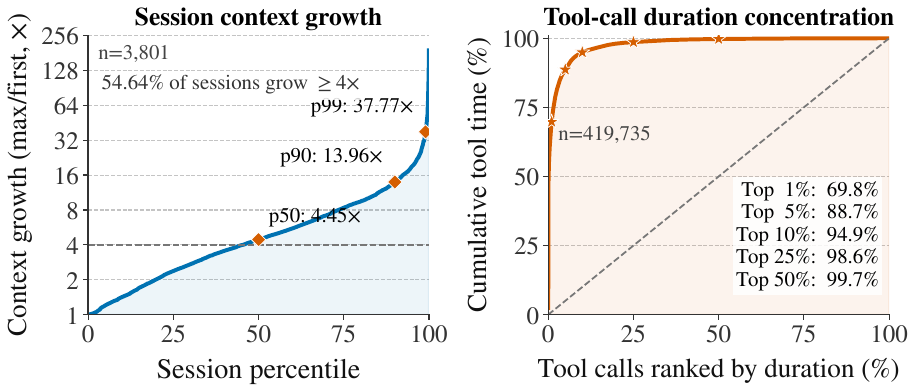}
    \caption{\textbf{\underline{Left:}} maximum-to-first context-growth ratio across session percentiles for
    3,801 multi-round sessions. \textbf{\underline{Right:}} cumulative share of total tool-call
    elapsed time contributed by 419,735 completed positive-duration tool calls
    ranked from longest to shortest. Dashed diagonal: uniform contribution.}
    \label{fig:char-session-tool}
\end{figure}

\noindent\textbf{Session context growth and tool-call durations vary widely.}
Across 3,801 multi-round sessions, the ratio between the maximum and
first-round context lengths is 4.45$\times$ at p50, 13.96$\times$ at p90,
and 37.77$\times$ at p99; 54.64\% of sessions grow by at least
4$\times$. Tool-call duration also varies widely. Among 419,735 completed tool calls, duration is 0.209~s at p50, 8.866~s at p90,
30.013~s at p95, and 180.002~s at p99.
\mbox{\autoref{fig:char-session-tool}} summarizes these session-growth and tool-return
distributions.

\begin{findingbox}
\textbf{\underline{Finding 4.}} Uneven context growth and widely varying tool-call
durations cause coding-agent sessions to reach different \revadd{context lengths} and
resume execution asynchronously.
\end{findingbox}

\begin{figure}[t]
    \centering
    \includegraphics[width=\linewidth]{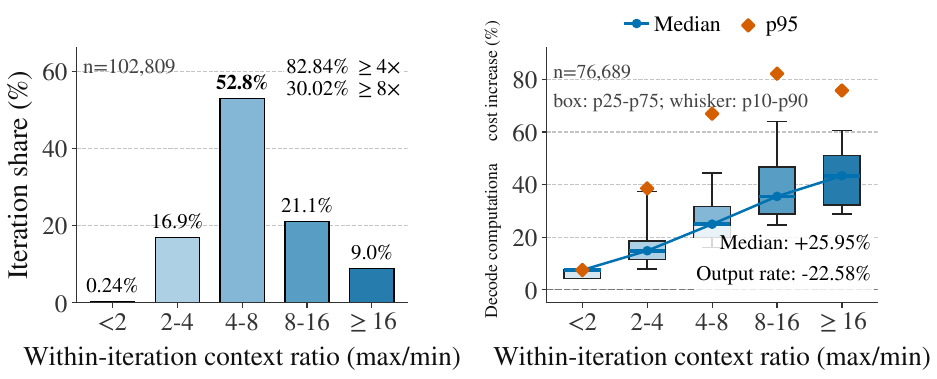}
    \caption{\textbf{\underline{Left:}} share of multi-request decode iterations grouped by their
    within-iteration \revadd{context-length ratio}, defined as the maximum divided by the
    minimum materialized context length. \textbf{\underline{Right:}} decode computational cost increase relative
    to a near-uniform context composition with the same request count and total
    context tokens, grouped by the observed \revadd{context-length ratio}. Boxes span p25 to
    p75, center lines mark the median, whiskers span p10 to p90, and diamonds
    mark p95.}
    \label{fig:char-context-cost}
\end{figure}

\noindent\textbf{Context lengths vary widely within decode iterations.}
We examine each context vector immediately after the standard continuous
batching planner forms a decode iteration. Across 102,809 multi-request
iterations, the largest materialized context is at least 2$\times$ the
smallest in 99.76\% of iterations, at least 4$\times$ in 82.84\%, and at
least 8$\times$ in 30.02\%. The largest-to-smallest \revadd{context-length ratio} is
6.32$\times$ at p50 and 19.69$\times$ at p95, while the context-length CV
is 0.58 at p50 and 0.93 at p95. The left panel of
\mbox{\autoref{fig:char-context-cost}} summarizes this within-iteration heterogeneity.

\begin{findingbox}
\textbf{\underline{Finding 5.}} Standard continuous batching frequently places
coding-agent requests with widely different materialized context lengths in
the same decode iteration.
\end{findingbox}

\noindent\textbf{High prefix reuse makes coding-agent execution
decode-dominant.}
Tool-follow-up requests account for 87.89\% of requests in the trace and
achieve 97.55\% token-weighted prefix reuse. Consequently, returning sessions
largely reuse their accumulated prefixes and prefill only newly appended
tokens. During the stable trace replay, 752,143 of 754,670 model iterations,
or 99.67\%, contain only decode work. These pure-decode iterations account for
79.34\% of aggregate model-execution time, compared with 13.69\% for mixed
prefill-decode iterations and 6.97\% for prefill-only iterations.

\begin{findingbox}
\textbf{\underline{Finding 6.}} High prefix reuse makes coding-agent execution
decode-dominant, with pure-decode iterations consuming 79.34\% of
model-execution time.
\end{findingbox}

\noindent\textbf{Within-batch context heterogeneity reduces decode efficiency.}
We analyze 76,689 pure-decode iterations from the trace-replay. For each
iteration, we compare its observed context-length composition with a
near-uniform composition containing the same number of requests and exactly
the same total context tokens. As shown in the right panel of
\mbox{\autoref{fig:char-context-cost}}, the observed composition has a 25.95\%
higher median computational cost per iteration, and 95.84\% of the iterations incur at least
10\% higher computational cost. Across the same output-token work, the output rate
decreases by 22.58\%, from 320.791 to 248.368 tokens/s.

\begin{findingbox}
\textbf{\underline{Finding 7.}} Decode efficiency depends on within-batch context
composition even when request count and total context-token volume remain
unchanged.
\end{findingbox}

\noindent\textbf{Decode-context composition differs across workers.}
We replay multi-worker serving with SMetric's KV-reuse- and load-aware routing~\cite{wang2026smetric}.
At each routing decision, we record SMetric's selected worker and the other
workers that satisfy the same KV-affinity and load constraints, and measure the
context-length distance on each worker. Among 12,160 routing decisions, 3,135,
or 25.78\%, retain at least two candidate workers after these constraints. In
2,444 of these decisions, or 77.96\%, the most context-compatible candidate
differs from SMetric's selected worker. Selecting that candidate reduces the
mean context-length distance by 34.25\%, showing frequent opportunities to
reduce \revadd{mixed-context interference} identified in \textbf{Finding~7}.

\begin{findingbox}
\textbf{\underline{Finding 8.}} Even after preserving KV reuse and balancing load, workers can have different decode-context compositions, making
\revadd{context-length compatibility} a meaningful routing signal.
\end{findingbox}

\section{System Design: The \sys Admission Layer}
\label{sec:system-design-1}

\sys introduces an admission layer between logical request readiness and
physical GPU execution. The layer maintains three complementary readiness
views. (\textbf{\underline{i}}) \textit{Logical readiness} indicates whether
a request is \revadd{eligible to proceed}. (\textbf{\underline{ii}}) \textit{State
readiness} characterizes the work required to make a request's KV state
available for GPU execution. (\textbf{\underline{iii}}) \textit{Execution
readiness} characterizes whether
a request selected by state admission can be efficiently integrated into the
current model iteration. \sys bridges the gaps between these
readiness views through controlled admission.

\subsection{Admission Model}
\label{subsec:admission-model}

We define three readiness views applied in sequence along the path from
request availability to GPU execution:
\begin{itemize}[leftmargin=*]
    \item \textbf{Logical readiness.} Logical readiness characterizes whether a
    request is \revadd{eligible to proceed}. A request becomes logically ready when
    (\textbf{\underline{i}}) a user submits a new request or
    (\textbf{\underline{ii}}) a tool call returns and the agent session can
    continue.

    \item \textbf{State readiness.} State readiness characterizes the work
    required to make a request's KV state available in GPU HBM. The three cases
    follow the storage hierarchy from highest to lowest readiness:
    (\textbf{\underline{i}}) the KV state is resident in HBM
    and immediately usable; (\textbf{\underline{ii}}) reusable KV state resides
    in CPU DRAM and requires restoration; or
    (\textbf{\underline{iii}}) reusable KV state exists only on SSD and must be
    loaded, or no reusable copy exists and the prefix must be recomputed.

    \item \textbf{Execution readiness.} Execution readiness characterizes the
    \revadd{context-length compatibility} between a state-selected request and the \revadd{current participating requests}. It distinguishes two conditions:
    (\textbf{\underline{i}}) the request is compatible with the active context
    composition and can execute efficiently with the existing population; or
    (\textbf{\underline{ii}}) the request introduces substantial context
    heterogeneity, requiring the resulting population to be reorganized for
    efficient model execution.
\end{itemize}

\begin{figure}
    \centering
    \includegraphics[width=0.95\linewidth]{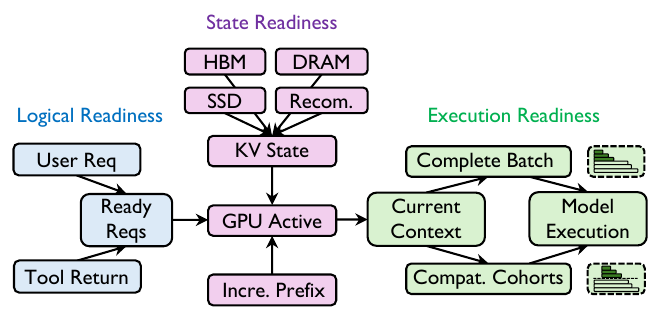}
    \caption{\textbf{Illustration of the three readiness stages in a request's
lifecycle.} Logical readiness indicates that a request is available after a
user request arrives or a tool call returns. State readiness characterizes the
work required to make its KV state available in GPU HBM, including tier
restoration, recomputation, and incremental-input prefill. Execution readiness
indicates whether the request can execute efficiently with the currently active
requests: compatible contexts execute as one complete batch, while
heterogeneous contexts are organized into compatible cohorts (detailed in~\S\ref{subsec:compat-aware-execution-admission}).}
    \label{fig:three-readiness}
\end{figure}

Together, these views define the admission path: the admission layer uses
logical readiness to form the candidate set, state readiness to \revadd{select which request is prepared next}, and execution readiness to organize \revadd{the participating requests} for efficient \revadd{attention execution}.
\autoref{fig:three-readiness} illustrates this three-stage admission path.

\noindent\textbf{Dynamic readiness.} Readiness evolves rapidly in coding-agent
serving. User requests and asynchronous tool returns continuously change the
logically ready population; long-lived session KV states grow and move across
storage tiers; and resumed sessions repeatedly reshape the active context
composition (\textbf{Findings~1-5}). Consequently, state and execution readiness may change after a
request first becomes logically ready. \sys reevaluates these views at their
corresponding decision opportunities rather than fixing them at request
arrival.

\noindent\textbf{Admission decisions.} To bridge these dynamically changing
readiness \revadd{views}, the admission layer makes two decisions:
(\textbf{\underline{i}}) \textit{State admission} determines which logically
ready request \revadd{has its KV state prepared next} according to its state readiness.
(\textbf{\underline{ii}}) \textit{Execution admission} determines how the
\revadd{participating requests are} organized for \revadd{attention execution} according to
\revadd{their} execution readiness. The following subsections describe Tiered-Aging \revadd{state admission} (\S\ref{subsec:tiered-aging-state-admission}) and Compatibility-Aware
\revadd{execution admission} (\S\ref{subsec:compat-aware-execution-admission}).

\subsection{Tiered-Aging State Admission}
\label{subsec:tiered-aging-state-admission}

\noindent\textbf{Admission challenge.} State admission selects which logically
ready request \revadd{has its KV state prepared next}. Long-lived coding-agent sessions
accumulate large reusable prefixes that may move across storage tiers during
tool-induced idle periods, while asynchronous tool returns continuously reshape
the ready population. Consequently, newly ready requests incur substantially
different computational costs for state preparation (\textbf{Findings~1-4}). \sys uses Tiered-Aging \revadd{state admission} to
exploit these differences while preserving progress.

\noindent\textbf{Tier-aware computational cost of state preparation.} As described in
\S\ref{subsec:admission-model}, the current placement of a request's KV state
determines the work required before GPU execution. A coding-agent continuation
contains a persistent session prefix and an incremental suffix introduced by a
user request or tool return. The reusable prefix may be distributed across
storage tiers, while the incremental suffix requires prefill computation. At
decision time $\text{t}$, \sys estimates the computational cost of state preparation
for each request $\text{i}$:
\begin{equation}
\text{C}_\text{i}^{\mathrm{state}}(\text{t}) =
\text{T}_\text{i}^{\mathrm{restore}}(\text{t})
+ \text{T}_\text{i}^{\mathrm{load}}(\text{t})
+ \text{T}_\text{i}^{\mathrm{prefill}}(\text{t}),
\end{equation}
where the three terms capture CPU-to-GPU restoration, SSD-to-HBM loading, and
prefill computation for incremental suffix tokens or unavailable prefix tokens,
respectively.
HBM-resident portions contribute no computational cost for restoration or loading, while
different portions of the same KV state may contribute to different terms.
Let $\mathcal{F}(\text{t})$ denote the logically ready requests that satisfy
the current HBM-capacity constraint; all other requests are temporarily
infeasible. If $\mathcal{F}(\text{t})=\varnothing$, no state-admission decision
is made.

\noindent\textbf{State computational cost preference.} Using the tier-aware computational cost,
\sys prefers the feasible request requiring the least work to make its KV state
HBM-ready:
\begin{equation}
\text{i}_{\mathrm{SCF}}(\text{t}) =
\arg\min_{\text{i}\in\mathcal{F}(\text{t})} \text{C}_\text{i}^{\mathrm{state}}(\text{t}).
\end{equation}
This preference applies the shortest-job-first principle, which prioritizes
shorter jobs to reduce aggregate waiting~\cite{smith1956various}. Here,
``job size'' refers to state-preparation work rather than the request's total
execution time. For coding-agent workloads, prioritizing shorter preparation
paths prevents a returning session with a large lower-tier KV state from
blocking multiple HBM-ready session continuations.

\noindent\textbf{Age-adjusted preference.} Selecting
$\text{i}_{\mathrm{SCF}}(\text{t})$ improves immediate efficiency, but
asynchronous tool returns may continuously introduce computationally inexpensive HBM-ready
continuations and repeatedly bypass sessions whose growing KV states reside in
lower tiers. Let
$\text{r}_\text{i}^{\mathrm{ready}}$ denote the most recent time at which a
user-request arrival or tool completion made request $\text{i}$ logically
ready. Its ready-wait age is
$\text{w}_\text{i}(\text{t})=\text{t}-\text{r}_\text{i}^{\mathrm{ready}}$.
Time spent executing the preceding tool is excluded because the request is not
yet logically ready. \sys uses this age to adjust the state computational cost:
\begin{equation}
\widehat{\text{C}}_\text{i}(\text{t}) =
\frac{\text{C}_\text{i}^{\mathrm{state}}(\text{t})}
     {1+\text{w}_\text{i}(\text{t})/\tau},
\end{equation}
where $\tau>0$ controls how quickly waiting weakens the state computational cost preference.\footnote{\label{fn:state-calibration}The aging timescale $\tau$ can be calibrated
offline on representative workloads to balance \revadd{SLO-compliant throughput} and admission
waiting, while $\text{W}_{\max}$ can be configured according to the evaluation configuration's
admission-waiting tolerance. Both remain fixed across evaluation workloads;
the settings are reported
in~\autoref{appendix:profiling}.}
Requests with lower computational costs of state preparation are therefore preferred initially, while
accumulated age progressively increases the competitiveness of requests with
more computationally expensive state preparation.

\begin{figure}
    \centering
    \includegraphics[width=0.95\linewidth]{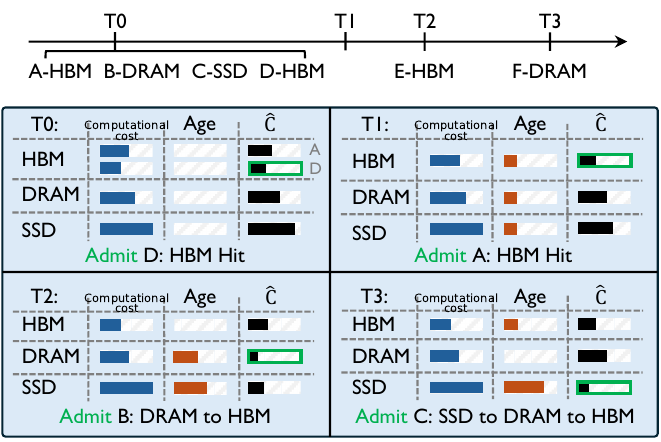}
    \caption{\textbf{Illustrative example of \revadd{Tiered-Aging} state admission.} A-HBM means that Request A arrives with its KV state resident in HBM. At successive admission opportunities, requests with HBM-resident KV state
and less additional prefill work are initially preferred because they incur
lower computational costs for state preparation, while aging progressively increases the
admission preference of requests that remain ready.}
    \label{fig:ta}
\end{figure}

\noindent\textbf{Progress-bounded selection.} Age adjustment progressively
relaxes the state computational cost preference, while a final progress guard strictly
bounds how long feasible requests can be repeatedly bypassed. Let
$\text{o}(\text{t})=\arg\max_{\text{i}\in\mathcal{F}(\text{t})} \text{w}_\text{i}(\text{t})$
denote the oldest feasible request. \sys makes the Tiered-Aging decision as
follows:
\begin{equation}
\text{i}^{*}(\text{t})=
\begin{cases}
\text{o}(\text{t}), & \text{w}_{\text{o}(\text{t})}(\text{t}) \geq \text{W}_{\max},\\[1mm]
\displaystyle\arg\min_{\text{i}\in\mathcal{F}(\text{t})}
\widehat{\text{C}}_\text{i}(\text{t}), & \text{otherwise},
\end{cases}
\end{equation}
where $\text{W}_{\max}>0$ bounds how long the state-efficiency preference may
override ready-time order.\textsuperscript{\ref{fn:state-calibration}}
Below this bound, \sys selects the request with the lowest age-adjusted state
computational cost. Once the bound is reached, the oldest feasible request takes precedence.
The guard overrides the efficiency preference but never violates the HBM
feasibility constraint.
\autoref{fig:ta} illustrates how the computational cost of state preparation and waiting age affect
admission order.

\noindent\textbf{Dynamic reevaluation.} At every state-admission opportunity,
\sys recomputes $\mathcal{F}(\text{t})$,
$\text{C}_\text{i}^{\mathrm{state}}(\text{t})$, and
$\text{w}_\text{i}(\text{t})$ while retaining each request's ready timestamp
$\text{r}_\text{i}^{\mathrm{ready}}$. User-request arrivals and tool
completions immediately introduce new candidates, while tier movement during
tool-induced idle periods changes their state computational costs. Tier-aware computational cost therefore
captures where a
persistent session's KV state resides when the session resumes, while age
captures how long the continuation has waited after becoming logically ready.

\noindent\textbf{State admission outcome.} After selecting
$\text{i}^{*}(\text{t})$, \sys performs the state-preparation work indicated by
the request's current state readiness, combining HBM reuse, CPU restoration,
SSD loading, and prefix recomputation as needed.
Once the required KV state becomes available in HBM, the request \revadd{proceeds} to execution admission. State admission therefore
determines \emph{which} request is prepared next, while execution admission
determines \emph{how} \revadd{the participating requests execute}.

\subsection{Compatibility-Aware Execution Admission}
\label{subsec:compat-aware-execution-admission}

\noindent\textbf{Admission challenge.} Execution admission determines how the
\revadd{requests} participating in a model iteration are organized for
\revadd{attention execution}. The contexts of long-lived coding-agent sessions grow at different
rates, while the sessions resume asynchronously after tool calls. Consequently,
requests with substantially different materialized context lengths frequently
execute together, reducing decode efficiency (\textbf{Findings~4-7}). \sys therefore uses
Compatibility-Aware \revadd{execution admission} to select an efficient execution plan
based on the current context composition.

\noindent\textbf{Context-length view.} At an eligible pure-decode iteration
occurring at time $\text{t}$, let $\mathcal{D}(\text{t})$ denote the
participating requests, $\text{n}(\text{t})=|\mathcal{D}(\text{t})|$, and
$\text{L}_\text{i}(\text{t})$ the materialized context length of request
$\text{i}$. Let $\pi_\text{j}$,
$\text{j}=1,\ldots,\text{n}(\text{t})$, be the request ranked $\text{j}$-th by
context length, so that
\begin{equation}
\text{L}_{\pi_1}(\text{t}) \leq
\text{L}_{\pi_2}(\text{t}) \leq \cdots \leq
\text{L}_{\pi_{\text{n}(\text{t})}}(\text{t}).
\end{equation}
These context lengths are obtained directly from the current execution state
rather than from predictions of generation length or future tool-return time.
This allows \sys to observe the context heterogeneity created by coding-agent
sessions at different stages of context growth.

\noindent\textbf{Compatibility-aware cohort construction.} To evaluate whether
cohorting is beneficial, each candidate cut $\text{k}$ partitions the
context-ordered requests into a shorter-context cohort and a longer-context
cohort:
\begin{equation}
\mathcal{D}_{\text{k}}^{\mathrm{short}}(\text{t})
=
\{\pi_1,\ldots,\pi_{\text{k}}\},
\quad
\mathcal{D}_{\text{k}}^{\mathrm{long}}(\text{t})
=
\{\pi_{\text{k}+1},\ldots,\pi_{\text{n}(\text{t})}\}.
\end{equation}
To avoid inefficiently small cohorts, each cohort must contain at least
$\text{m}(\text{t})=\max(\text{M}_{\min},
\lceil\rho\,\text{n}(\text{t})\rceil)$ requests. Here,
$\text{M}_{\min}\in\mathbb{N}^{+}$ is the minimum efficient cohort size, and
$0<\rho\leq 1/2$ prevents highly imbalanced splits.\footnote{\label{fn:execution-calibration}The minimum cohort size $\text{M}_{\min}$, minimum cohort fraction $\rho$,
gain threshold $\Gamma$, and computational cost model for attention execution are calibrated offline once
for each model-hardware configuration and remain fixed across workloads, as detailed
in~\autoref{appendix:profiling}.}
Accordingly,
$\mathcal{K}(\text{t})$ contains the integer cuts satisfying
$\text{m}(\text{t})\leq\text{k}\leq
\text{n}(\text{t})-\text{m}(\text{t})$. If
$\text{n}(\text{t})<2\text{m}(\text{t})$, this set is empty, and \sys retains
complete-population execution.

\noindent\textbf{Compatibility-aware preference.} When
$\mathcal{K}(\text{t})\neq\varnothing$, \sys compares each feasible cut's
estimated computational cost of two-cohort attention with that of complete-population attention
and selects the cut with the largest predicted relative gain:
\begin{equation}
\text{k}^{*}(\text{t})
=
\arg\max_{\text{k}\in\mathcal{K}(\text{t})}
\frac{
\widehat{\text{C}}^{\mathrm{full}}(\text{t})
-
\widehat{\text{C}}_{\text{k}}^{\mathrm{cohort}}(\text{t})
}{
\widehat{\text{C}}^{\mathrm{full}}(\text{t})
}.
\end{equation}
Here, $\widehat{\text{C}}^{\mathrm{full}}(\text{t})$ denotes the estimated
attention-execution time of the complete population, while
$\widehat{\text{C}}_{\text{k}}^{\mathrm{cohort}}(\text{t})$ denotes the
combined attention-execution time of the two cohorts, including their
additional computational overhead. The two-cohort organization is applied only
when its conservative predicted relative gain is at least $\Gamma=0.05$;
otherwise, \sys retains complete-population execution.\textsuperscript{\ref{fn:execution-calibration}}

\begin{figure}
    \centering
    \includegraphics[width=0.95\linewidth]{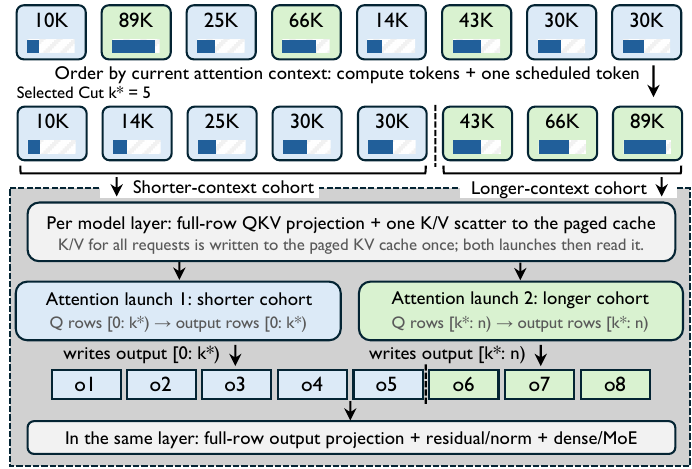}
    \caption{\textbf{Illustrative example of Compatibility-Aware execution
admission.} \sys reorders a heterogeneous pure-decode batch by context length
and partitions it into shorter- and longer-context cohorts. Each model layer
computes QKV and updates the shared paged KV cache once, after which the two
cohorts execute through separate attention launches and write to disjoint
regions of the same output buffer. All requests remain in the iteration, and
the remaining layer computation proceeds over the complete reordered batch.}
    \label{fig:compat-aware}
\end{figure}

\noindent\textbf{Compatibility-aware execution.} Under the selected cohort
organization, the model performs a single shared dense/MoE traversal and
preserves normal KV-update semantics while executing attention in two
context-compatible launches. Both cohorts execute within the same model
iteration and together cover exactly the requests in
$\mathcal{D}(\text{t})$. Thus, execution admission changes the physical
organization of attention without duplicating complete model execution or
changing the participating request set.
\autoref{fig:compat-aware} illustrates this cohort-based organization of
attention execution.

\noindent\textbf{Dynamic reevaluation.} At every eligible pure-decode
iteration, \sys reconstructs $\mathcal{D}(\text{t})$ and reevaluates its
materialized context composition. Session progress, request completion, and
asynchronous tool returns may therefore change the preferred organization from
one iteration to the next. \sys falls back to complete-population execution
when the population is insufficiently heterogeneous, the execution shape is
unsupported, or the predicted gain is insufficient.

\section{System Design: The \sys Routing Layer}
\label{sec:system-design-2}

For coding-agent serving with multiple workers, \sys introduces a routing layer
that determines the worker assignment for each arriving request. The layer
considers three complementary placement signals.
(\textbf{\underline{i}}) \textit{KV reuse} captures the reusable session prefix
available at each worker. (\textbf{\underline{ii}}) \textit{Worker load}
captures the pressure created by active requests.
(\textbf{\underline{iii}}) \textit{\revadd{Context-length compatibility}} captures how closely
the arriving context matches each worker's active decode population. \sys
applies these signals hierarchically: it preserves useful KV affinity when
load-safe and uses \revadd{context-length compatibility} to select among the remaining
load-eligible workers (\textbf{Finding~8}).

\noindent\textbf{Reuse- and load-aware placement.} \sys adopts SMetric's
reuse- and load-aware policy~\cite{wang2026smetric} as its parent routing
policy. A session's first request initially selects a least-loaded worker. For
a session continuation, the router prefers the worker with the largest actual
local-prefix hit and preserves this affinity while the worker remains within
the configured overload bound. This policy exploits the locality of long-lived
coding-agent sessions while limiting load imbalance caused by asynchronous
tool-return bursts.

\begin{figure}
    \centering
    \includegraphics[width=0.95\linewidth]{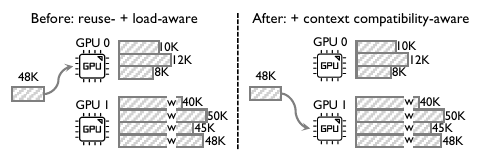}
    \caption{\textbf{Illustrative example comparing routing decisions.}
When KV affinity is inconclusive, reuse- and load-aware routing selects the
minimum-load GPU, while \sys selects a load-safe GPU whose active contexts
better match the arriving 48K request.}
    \label{fig:routing-pic}
\end{figure}

\noindent\textbf{Compatibility-aware placement.} When KV affinity does not determine the destination, \sys considers workers at
the minimum load or one active request above that minimum, and selects among
them according to \revadd{context-length compatibility}. For an arriving request $\text{i}$ at routing
time $\text{t}$, its compatibility with worker $\text{w}$ is:
\begin{equation}
\text{d}_{\text{i},\text{w}}(\text{t})
=
\frac{1}{|\mathcal{C}_{\text{w}}(\text{t})|}
\sum_{\ell\in\mathcal{C}_{\text{w}}(\text{t})}
\left|
\log_2
\frac{\ell}{\text{L}_{\text{i}}^{\mathrm{arr}}(\text{t})}
\right|.
\end{equation}
Here, $\mathcal{C}_{\text{w}}(\text{t})$ is the multiset of materialized
context lengths of worker $\text{w}$'s active decoders,
$|\mathcal{C}_{\text{w}}(\text{t})|$ is the number of those decoders,
$\ell$ is one decoder's context length, and
$\text{L}_{\text{i}}^{\mathrm{arr}}(\text{t})$ is the arriving request's
reusable prefix plus incremental input. The resulting
$\text{d}_{\text{i},\text{w}}(\text{t})$ is a dimensionless routing
\revadd{context-length distance}; a smaller value indicates a more compatible worker.
\sys selects the eligible worker with the smallest distance, forming more
homogeneous decode populations while preserving the parent policy's reuse and
load protections.
\autoref{fig:routing-pic} illustrates how \revadd{context-length compatibility} can change the
routing decision when KV affinity is inconclusive.

\noindent\textbf{Dynamic reevaluation.} \sys refreshes local KV reuse, worker
load, and active context composition at every routing opportunity. These
measurements reflect only the current system state and do not rely on predicted
output lengths or future tool-return times. When context telemetry is
unavailable or no load-eligible worker has an active decode population, \sys
retains the reuse- and load-aware destination. Routing changes only the arriving
request's assignment and does not migrate requests that are already GPU-active.
Once assigned, the request enters the selected worker's admission layer
(\S\ref{sec:system-design-1}) for KV-state preparation and model execution.

\section{Implementation}
\label{sec:implementation}

\noindent\textbf{Runtime integration.} We implement \sys as an orchestration
layer with approximately 2.9\,KLoC of core policy and planning code, connected
to the inference backend through an adapter. The adapter adds hooks to request
dispatch, waiting-queue selection, and model execution. At each hook, \sys
reads the current runtime state and returns a routing, state-admission, or
execution-admission decision. The backend continues to handle memory
allocation, KV-state movement, preemption, model execution, and sampling.

\noindent\textbf{Routing and admission paths.} The dispatcher maintains
asynchronously refreshed worker snapshots and uses a consistent snapshot
version for each arriving request. The resulting routing decision is attached
to the request before normal dispatch. At each state-admission opportunity,
\sys examines a bounded frontier of at most 64 candidate entries and queries
the KV manager for reusable-prefix matches, KV-block demand, and usable HBM
capacity. Candidate inspection is read-only, and the selected entry is promoted
through one remove-and-prepend operation before native scheduling continues.
For an accepted execution plan, \sys reorders the persistent batch through
state-safe row swaps and constructs cohort-specific views of the required
attention metadata. Each model layer performs one complete-population KV
update, invokes the existing attention operator once for each
context-compatible cohort, and writes each cohort's results into the
corresponding region of the original output buffer. Dense or MoE computation
and sampling remain unchanged.

\begin{figure}
    \centering
    \includegraphics[width=0.95\linewidth]{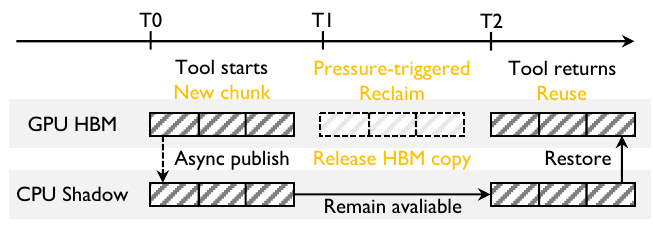}
    \caption{\textbf{Illustrative example of CPU-shadow offloading.}
A completed KV-state chunk is asynchronously published from GPU HBM to the
CPU shadow. Under HBM pressure, the GPU copy can be reclaimed without an
eviction-time transfer, while the CPU copy remains available for subsequent
restoration and reuse.}
    \label{fig:cpu-shadow}
\end{figure}

\noindent\textbf{CPU-shadow offloading and HBM reclaim.} The KV connector
asynchronously publishes completed KV-state chunks to a shared CPU shadow in
DRAM and restores matching KV-state chunks on demand. Once publication
completes, the corresponding HBM copy can be reclaimed without an eviction-time
GPU-to-CPU transfer. \autoref{fig:cpu-shadow} illustrates CPU-shadow
publication, HBM reclaim, and on-demand restoration.
When decode execution requires additional HBM blocks, the
backend automatically invokes HBM reclaim. Following CacheWise's cache-reclaim
integration~\cite{tiwari2026cachewise}, \sys determines the HBM reclaim victims
using a tool-call-aware policy. Within a bounded victim window, it prioritizes
valid, uniquely owned KV states predicted to remain inactive longer based on
tool-call identity and elapsed tool-call time. Incomplete, shared, or
unverified KV states retain their native LRU order.

\section{Evaluation}
\label{sec:evaluation}

We design the evaluation\footnote{All research reported in this paper, including
access to and use of models, datasets, and code, was conducted by the
university-affiliated authors.} to answer two research questions:

\begin{itemize}[leftmargin=*]
    \item \textbf{RQ1:} \textit{How does \sys compare with existing
    systems in throughput, SLO-compliant throughput, and latency across
    workloads and model sizes?}
    \item \textbf{RQ2:} \textit{How do routing, state
    admission, and execution admission contribute to \sys's performance,
    and what runtime behavior explains their benefits?}
\end{itemize}

\subsection{Experimental Setup}

\textbf{Hardware Environment.} We consider four configurations: (\textbf{\underline{i}}) Setup 1: One NVIDIA A100-80GB GPU for admission-only ablation. (\textbf{\underline{ii}}) Setup 2: 4 NVIDIA A100-80G GPUs, NVLink at 300 GB/s intra-node. (\textbf{\underline{iii}}) Setup 3: 16 NVIDIA A100-80G GPUs across two nodes (8 GPUs/node), NVLink at 300 GB/s intra-node, InfiniBand at 50 GB/s inter-node.

\noindent \textbf{Baseline systems.} We compare \sys with three baseline systems. (\textbf{\underline{i}}) CacheWise~\cite{tiwari2026cachewise} improves KV-cache reuse through prefix-aware scheduling and predictive eviction guided by tool-call metadata. (\textbf{\underline{ii}}) SMetric~\cite{wang2026smetric} balances load and KV-cache reuse by routing each session's first request based on load and subsequent requests based on cache locality. (\textbf{\underline{iii}}) vLLM~\cite{kwon2023efficient} serves as an unoptimized baseline using round-robin routing and FIFO request admission.

\noindent \textbf{Workload traces.} We derive workload traces by subsampling TraceLab~\cite{zhu2026tracelab} at three request arrival rates, denoted Q1, Q2, and Q3, corresponding to 3, 5, and 7 requests/s. The subsampling procedure preserves the key long-tailed properties of the original traces, including the distribution of tool-call intervals. For single-GPU admission ablation, we use M1, M2, and M3, which contain approximately one quarter of the sessions in Q1, Q2, and Q3, respectively, retaining complete sessions with unchanged token counts and tool-call intervals.

\noindent \textbf{Models and metrics.} We use Qwen3-Coder-30B-A3B-FP8 and Qwen3-235B-A22B-FP8 as the served models; both are popular open-source models~\cite{yang2025qwen3}. Each Qwen3-30B replica is served on a single GPU, whereas each Qwen3-235B replica is served on eight GPUs. Following prior work~\cite{tiwari2026cachewise,wang2026smetric}, we evaluate overall throughput, SLO-compliant throughput, tail latencies for end-to-end (E2E) latency, time between tokens (TBT), and time to first token (TTFT), as well as prefix reuse rate and queue length.

\subsection{End-to-end Evaluation (RQ1)}

We first examine whether \sys improves both serving throughput and latency.
The subsequent ablations separate the contributions of routing, state
admission, and execution admission, while the decision-level measurements
examine how these \revadd{mechanisms} affect runtime behavior.

\begin{figure}[t!]
    \centering
    \includegraphics[width=\linewidth]{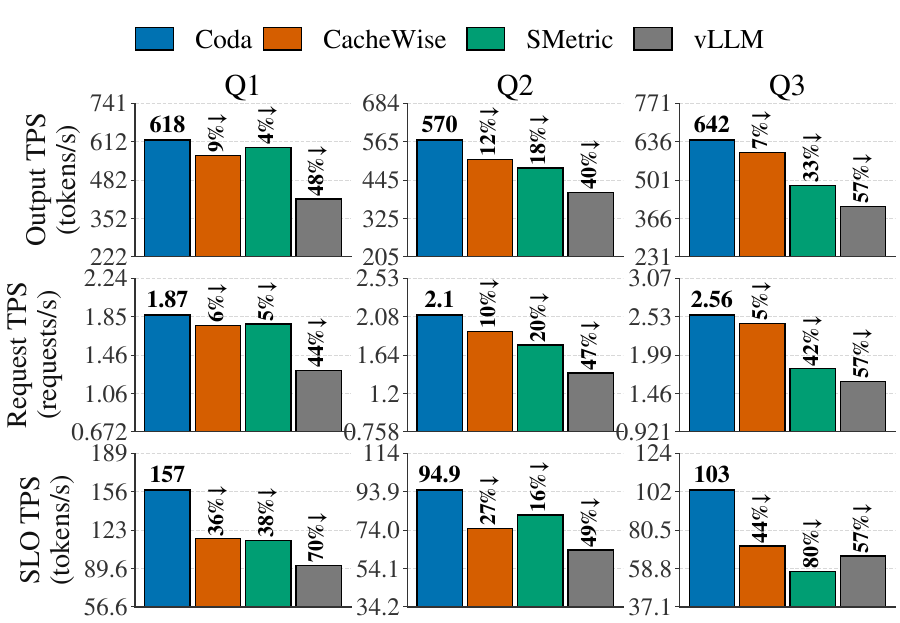}
    \caption{Throughput comparison for Qwen3-30B on Q1-Q3
    under setup~2: output, request, and SLO TPS.}
    \label{fig:30b-thpt}
\end{figure}

\noindent\textbf{30B end-to-end throughput.}
\sys achieves the highest \revadd{output-token}, completed-request, and SLO-compliant
throughput on all three workloads (\autoref{fig:30b-thpt}). Relative to the
strongest competing baseline for each metric and trace, the gains are
4.2-12.1\%, 4.9-10.0\%, and 15.9-44.3\%, respectively. The larger gains in
SLO-compliant throughput show that the improvement extends beyond raw token
generation to output that meets the latency targets. For example, on Q3,
\sys produces 642.1 output tokens/s and 103.2 SLO-compliant output tokens/s,
compared with CacheWise's 598.3 and 71.5. Relative to vLLM, the \revadd{output-token} and
SLO-compliant throughput gains reach 56.9\% and 70.3\%. The advantage over
CacheWise and SMetric further shows that the gains are not limited to replacing
the unoptimized baseline. We isolate the contributions of routing and
admission in \autoref{fig:ablation}.

\begin{figure}[t!]
    \centering
    \includegraphics[width=\linewidth]{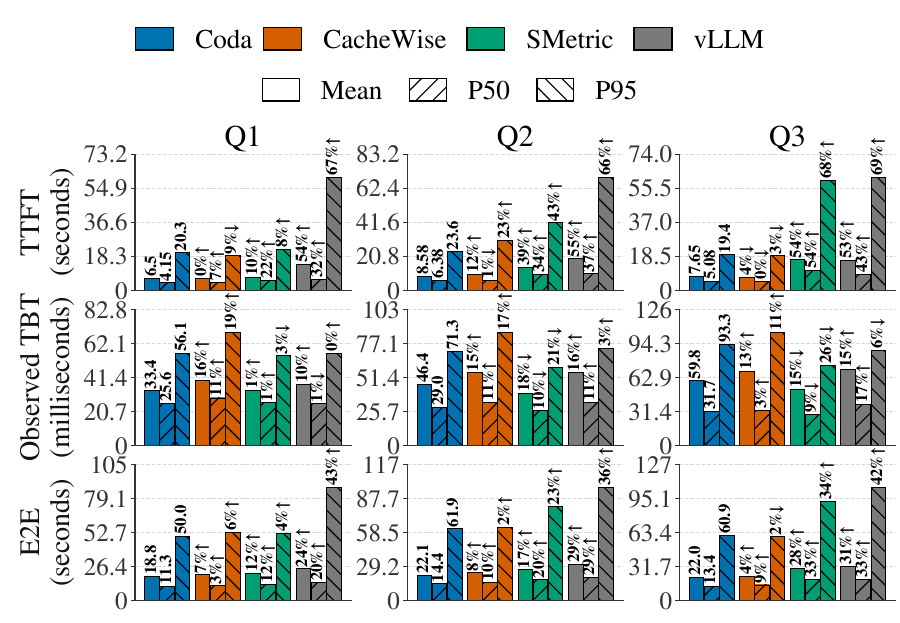}
    \caption{Latency comparison for Qwen3-30B on Q1-Q3 under setup~2: TTFT,
    observed TBT, and E2E latency.}
    \label{fig:30b-lat}
\end{figure}

\noindent\textbf{30B request latency (TTFT and E2E).}
\autoref{fig:30b-lat} reports TTFT and E2E latency, capturing request startup
and completion, respectively. Relative to SMetric, \sys reduces p50 and p95
TTFT by 22.0-53.5\% and 8.0-67.6\%, respectively. It also reduces p50 E2E
latency by 12.1-33.0\%. Its p95 E2E
latency is comparable on Q1 and 23.2-34.2\% lower on Q2 and Q3. Relative to
CacheWise, E2E latency is comparable or lower, with p50 reductions of
8.6-9.8\% on Q2 and Q3 and a p95 reduction of 5.9\% on Q1.
Compared with vLLM, \sys reduces p95 TTFT and E2E latency by 66.0-68.6\%
and 36.5-43.1\%, respectively. Taken together, these results show that \sys
improves request responsiveness
over SMetric and vLLM while maintaining comparable or lower completion
latency than CacheWise.

\noindent\textbf{30B token-delivery latency (TBT).}
\sys achieves comparable or lower TBT than CacheWise, including p95 reductions of
10.9-18.7\%. Relative to vLLM, p50 TBT is comparable on Q1 and 11.1-17.0\%
lower on Q2 and Q3; p95 TBT is comparable on Q1 and Q2 but 6.1\% higher on
Q3. Compared with SMetric, p50 and p95 TBT are comparable on Q1
but higher on Q2 and Q3. SMetric's lower TBT on these two traces is consistent
with smaller effective decode batches: the decode-concurrency in
\autoref{fig:cache_reuse_batch_size} is 13.0\% and 22.6\% higher for \sys.
In summary, \sys combines higher throughput with comparable or
lower TBT than CacheWise, while SMetric's lower TBT on Q2 and Q3 accompanies
lower throughput.

\begin{figure}[t!]
    \centering
    \includegraphics[width=\linewidth]{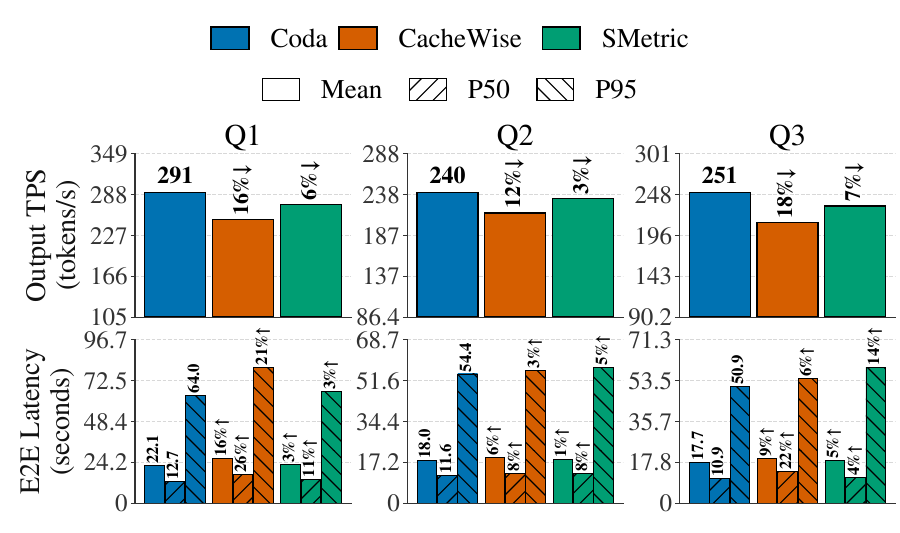}
    \caption{Throughput and E2E latency comparison for Qwen3-235B on Q1-Q3 under setup~3: output TPS and mean, p50, and p95 E2E latency.}
    \label{fig:235b}
\end{figure}

\noindent\textbf{235B results.}
\autoref{fig:235b} presents the attainable performance of \sys on the larger
model in a larger cluster. \sys achieves 11.7-18.3\% higher output-token
throughput than CacheWise and 3.1-7.2\% higher than SMetric. Its mean and
p50 E2E latency are 6.1-15.7\% and 8.2-25.9\% lower than CacheWise's,
respectively, while its p95 latency is comparable or up to 20.6\% lower.
Against SMetric, the mean E2E latency remains comparable, and the p50 and
p95 latencies are comparable or lower by up to 11.2\% and 14.3\%,
respectively. Overall, these comparisons demonstrate \sys's throughput
and E2E latency advantages for the 235B model.

\subsection{Ablation and Mechanism Analysis (RQ2)}

\begin{figure}[t!]
    \centering
    \includegraphics[width=\linewidth]{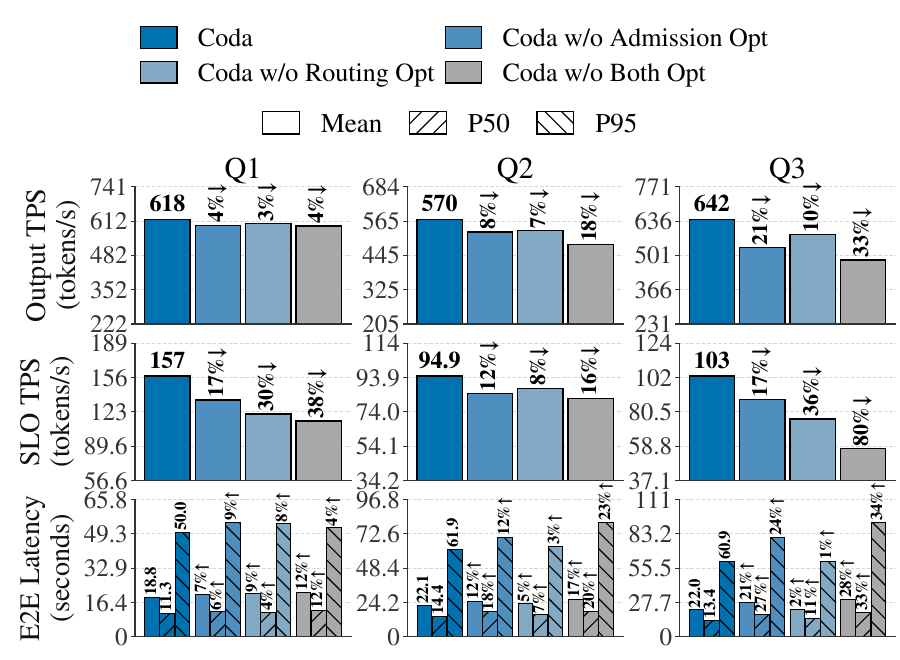}
    \caption{Routing and admission ablation for Qwen3-30B on Q1-Q3 under setup 2:
    output, SLO TPS, and E2E latency.}
    \label{fig:ablation}
\end{figure}

\noindent\textbf{Component ablation.}
\autoref{fig:ablation} isolates the contributions of routing and admission
by disabling routing optimization, admission optimization, or both. Compared
with the routing-only variant, the full system gains 3.7-20.8\% in
output-token throughput and 12.1-17.3\% in SLO-compliant throughput by
enabling admission. Compared with the admission-only variant, enabling
routing yields gains of 2.7-10.1\% and 8.2-35.7\% in these two metrics,
respectively. With both optimizations enabled, \sys outperforms the variant
with neither optimization by 4.2-33.2\% in output-token throughput and
15.8-80.3\% in SLO-compliant throughput. Its p50 E2E latency is also
12.1-33.0\% lower than this variant's, while p95 is comparable on Q1 and
23.2-34.2\% lower on Q2 and Q3. These gains support the complementary
roles of routing in shaping each worker's request and KV-state composition
and admission in controlling \revadd{preparation order and attention grouping}. Thus, the
combined design achieves the highest \revadd{output-token} and SLO-compliant throughput
across Q1-Q3 while maintaining comparable or lower E2E latency than every
ablated variant.

\begin{figure}[t]
    \centering
    \includegraphics[width=\linewidth]{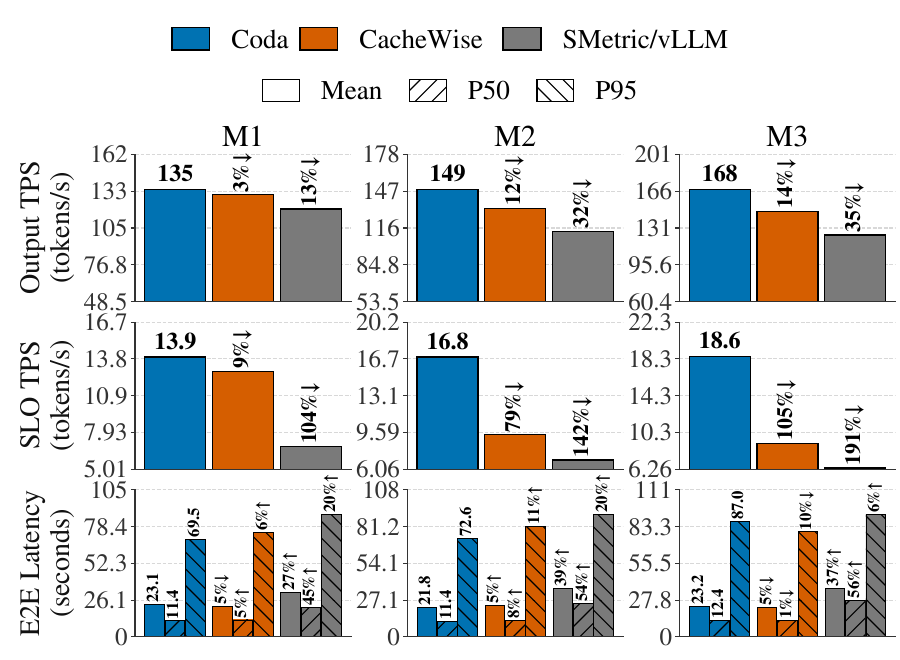}
    \caption{Performance comparison on M1-M3 under setup~1:
    output, SLO TPS, and mean, p50, and p95 E2E latency.}
    \label{fig:single-gpu}
\end{figure}

\noindent\textbf{Single-GPU admission ablation.}
We examine \sys's admission
performance in greater detail using single-GPU experiments
(\autoref{fig:single-gpu}). Compared with CacheWise, \sys
achieves 3.1-14.3\% higher output-token throughput and 8.8-104.7\% higher
SLO-compliant throughput, with both gains increasing from M1 to M3. Against
the shared SMetric/vLLM FIFO baseline, its output-token and SLO-compliant
throughput improve by 12.5-35.2\% and 104.4-190.7\%, respectively, while
mean, p50, and p95 E2E latency decrease by 27.3-39.0\%, 44.9-55.6\%, and
6.0-20.2\%. Compared with CacheWise, \sys maintains comparable E2E latency
on average across M1-M3; the equal-weight means of p50 and p95 latency are
3.8\% and 2.4\% lower, respectively. These results show that \sys's
admission layer increases \revadd{output-token} and SLO-compliant throughput while
maintaining comparable average E2E latency to CacheWise.

\begin{table}[t]
    \centering
    \caption{Cumulative ablation of single-GPU admission policies on M1-M3.
    TA applies Tiered-Aging state admission to the CacheWise baseline;
    CA further adds Compatibility-Aware execution admission.}
    \label{tab:single_gpu_admission_ablation}
    \resizebox{\linewidth}{!}{%
        \begin{tabular}{l | l | r | r | r}
            \hline
            \textbf{Workload} & \textbf{Policy} & \textbf{Output tok/s} &
            \textbf{Useful tok/s} & \textbf{SLO tok/s} \\
            \hline
            M1 & CacheWise & 130.792 & 1535.248 & 12.788 \\
               & +TA & 130.933 & 1542.755 & 12.370 \\
               & \textbf{+CA} & \textbf{134.800} & \textbf{1564.598} &
               \textbf{13.917} \\
            \hline
            M2 & CacheWise & 132.587 & 1784.415 & 9.378 \\
               & +TA & 143.473 & 1785.320 & 13.402 \\
               & \textbf{+CA} & \textbf{148.732} & \textbf{1819.360} &
               \textbf{16.830} \\
            \hline
            M3 & CacheWise & 146.622 & 2040.227 & 9.070 \\
               & +TA & 163.902 & 2085.903 & 16.883 \\
               & \textbf{+CA} & \textbf{167.652} & \textbf{2189.557} &
               \textbf{18.567} \\
            \hline
        \end{tabular}%
    }
\end{table}

\noindent\textbf{Admission-policy ablation.}
\autoref{tab:single_gpu_admission_ablation} examines two successive admission
refinements to the CacheWise baseline: Tiered-Aging state admission
(TA; \S\ref{subsec:tiered-aging-state-admission}), followed by
Compatibility-Aware execution admission
(CA; \S\ref{subsec:compat-aware-execution-admission}). In addition to
output-token and SLO-compliant throughput, we report useful throughput,
which counts each session's initial prefix once, appended input, and
in-window output for completed requests, divided by the window duration.
Compared with CacheWise, TA increases the equal-weight mean output-token,
useful, and SLO-compliant throughput across M1-M3 by 6.9\%, 1.0\%, and
36.6\%, respectively. Its \revadd{output-token} and SLO-compliant throughput are comparable
to CacheWise's on M1, with larger gains on M2 and M3. Adding CA while
keeping TA unchanged yields further gains of 2.9\%, 2.9\%, and 15.6\%
in these three mean throughput metrics, respectively. The additional
SLO-compliant throughput gain ranges from 10.0\% to 25.6\% across traces.
The combined policy achieves the highest values for all three metrics on
every trace, demonstrating the complementary benefits of state and execution
admission.

\begin{table}[t]
    \centering
    \caption{Changes in \sys's routing decisions relative to SMetric. Distances are averaged over changed
    decisions.}
    \label{tab:routing_decision_changes}
    \resizebox{\linewidth}{!}{%
        \begin{tabular}{l | r | r | r | r}
            \hline
            \textbf{Trace} & \textbf{Changes} & \textbf{SMetric Dist.} &
            \textbf{\sys Dist.} & \textbf{Reduction} \\
            \hline
            Q1 & 57 & 1.411 & 1.059 & 24.9\% \\
            \hline
            Q2 & 80 & 1.493 & 1.048 & 29.8\% \\
            \hline
            Q3 & 83 & 1.735 & 1.324 & 23.7\% \\
            \hline
            \textbf{Total/mean} & \textbf{220} & \textbf{1.563} &
            \textbf{1.155} & \textbf{26.1\%} \\
            \hline
        \end{tabular}%
    }
\end{table}

\noindent\textbf{Routing-decision changes.}
\autoref{tab:routing_decision_changes} examines how \revadd{context-length compatibility}
affects worker selection relative to SMetric's reuse- and load-aware routing.
During the five-minute measurement windows, \sys selects a different worker
in 57, 80, and 83 decisions on Q1, Q2, and Q3, respectively. For these
changed decisions, its selected workers reduce the mean \revadd{context-length distance}
score by 24.9\%, 29.8\%, and 23.7\% relative to SMetric's choices. Across
all 220 changed decisions, the pooled score decreases from 1.563 to 1.155,
a 26.1\% reduction. These results show that \sys's compatibility signal
changes placement toward closer \revadd{context-length matches}, complementing the
throughput benefits established by the routing ablation in
\autoref{fig:ablation}.

\begin{table}[t]
    \centering
    \caption{Admission-choice changes in \sys relative to CacheWise,
    with mean observed TBT for both systems.}
    \label{tab:kvm_admission_changes}
    \resizebox{\linewidth}{!}{%
        \begin{tabular}{l | r | r | r | r}
            \hline
            \textbf{Trace} & \textbf{Changes} & \textbf{CacheWise TBT} &
            \textbf{\sys TBT} & \textbf{Reduction} \\
            \hline
            M1 & 16 & 69.30 ms & 65.98 ms & 4.8\% \\
            \hline
            M2 & 37 & 81.79 ms & 69.47 ms & 15.1\% \\
            \hline
            M3 & 77 & 84.00 ms & 71.74 ms & 14.6\% \\
            \hline
            \textbf{Total/mean} & \textbf{130} & \textbf{78.36 ms} &
            \textbf{69.06 ms} & \textbf{11.9\%} \\
            \hline
        \end{tabular}%
    }
\end{table}

\noindent\textbf{Admission-choice changes.}
\autoref{tab:kvm_admission_changes} compares \sys's admission choices and
mean observed TBT with CacheWise's. Within the five-minute decision windows,
\sys makes 16, 37, and 77 different admission choices on M1, M2, and M3,
respectively. Over the 10-minute measurement windows, its mean TBT is comparable
to CacheWise's on M1 and 15.1\% and 14.6\% lower on M2 and M3. Averaging
the three trace-level means equally gives 69.06~ms for \sys versus
78.36~ms for CacheWise, an 11.9\% reduction. These TBT measurements describe
token delivery over each 10-minute window, not only requests with changed
admission choices.
Together with \autoref{tab:single_gpu_admission_ablation}, this comparison
shows that \sys changes admission behavior while achieving higher throughput
and lower average token-delivery latency.

\begin{figure}[t]
    \centering
    \includegraphics[width=\linewidth]{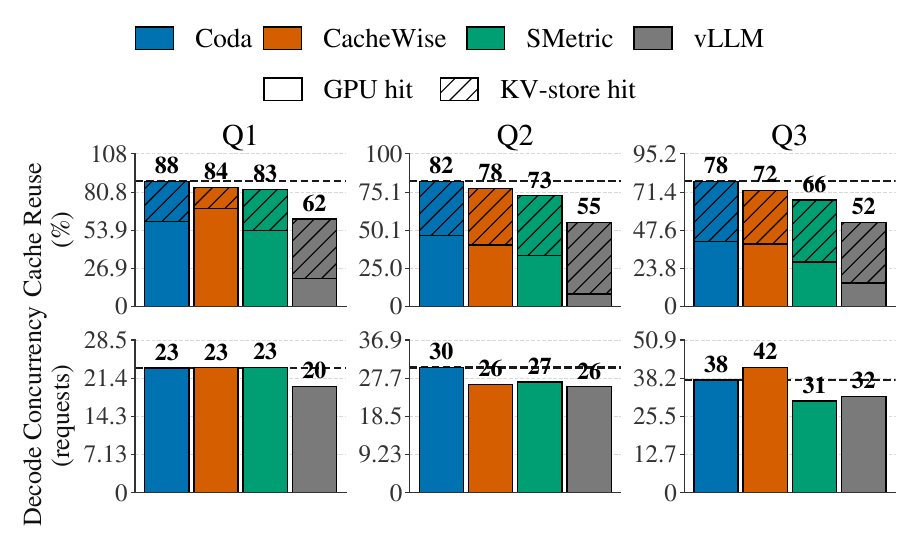}
    \caption{GPU and KV-store hit rates and the mean decode-concurrency on Q1-Q3 under setup 2.}
    \label{fig:cache_reuse_batch_size}
\end{figure}

\noindent\textbf{KV reuse.}
\autoref{fig:cache_reuse_batch_size} shows the contributions of
GPU-resident and KV-store hits to total reuse. \sys achieves total reuse
rates of 88.3\%, 82.1\%, and 78.0\% on Q1-Q3, respectively, exceeding
CacheWise by 4.3-5.6 percentage points,
SMetric by 5.6-11.6 points, and vLLM by 25.7-27.0 points. The source
breakdown also distinguishes total reuse from GPU residency: on Q1,
CacheWise has a larger GPU-hit share, but \sys achieves higher total reuse
through additional KV-store hits. \sys thus achieves higher total reuse
across all three traces by drawing on both GPU-resident KV state and the
KV store.

\noindent\textbf{Decoding concurrency.}
The lower panel of \autoref{fig:cache_reuse_batch_size} demonstrates cluster-wide
decoding concurrency. \sys yields 23.3, 30.2, and 37.5 concurrent requests on
Q1-Q3, respectively. Its concurrency is comparable to SMetric's on Q1 and
13.0\% and 22.6\% higher on Q2 and Q3, respectively. It is also
16.7-18.1\% higher than vLLM's. On Q3, however, CacheWise has a larger
concurrency but lower \revadd{output-token} and SLO-compliant throughput than \sys.
This comparison reinforces the importance of combining KV reuse with
efficient execution rather than maximizing decoding concurrency alone.

\section{Conclusion}
\label{sec:conclusion}

We presented \sys, a coding-agent serving system that exploits admission
flexibility to improve serving efficiency while preserving request
progress. Through direct trace analysis and replay, we
showed that ready requests can incur substantially different KV-state
computational costs for state preparation and execute inefficiently together because of
heterogeneous context lengths. Guided by these findings, \sys introduces
Tiered-Aging state admission, Compatibility-Aware execution admission, and a separate routing
layer for multi-worker configurations. Experiments across models and workloads
demonstrate gains in output-token and SLO-compliant throughput, while
component ablations identify the contributions of admission and routing.
These results highlight the benefit of treating admission as a first-class
optimization target.

\bibliographystyle{plain}
\bibliography{reference}

\appendix

\section{Profiling Details}
\label{appendix:profiling}

\noindent\textbf{Profile scope.} \sys profiles each
model--hardware--serving-engine configuration once and binds the resulting
artifacts to that configuration. All estimated times are represented in
milliseconds. Changing the model, GPU, KV representation, transfer topology,
or serving engine requires reprofiling; within one configuration, profiles and
policy parameters remain fixed across workloads.

\noindent\textbf{Computational cost calibration for state preparation.} We profile the operations that
contribute to $\text{C}_\text{i}^{\mathrm{state}}(\text{t})$ separately.
CPU-to-GPU restoration is measured over multiple transfer sizes and represented
by a piecewise-linear H2D service-time curve. At runtime,
$\text{T}_\text{i}^{\mathrm{restore}}(\text{t})$ combines the profiled time for
the request's CPU-resident KV bytes with the outstanding completion time in its
transfer domain. Prefill is profiled as a function of the number of tokens that
lack reusable KV state; $\text{T}_\text{i}^{\mathrm{prefill}}(\text{t})$ is
obtained from the unavailable-prefix and incremental-suffix token counts. The
last measured segment is used for positive-slope extrapolation beyond either
profile's largest point. HBM-resident portions incur no computational cost for restoration,
and the evaluated CPU-shadow configuration incurs no eviction-time D2H transfer.
The current prototype and reported experiments do not configure an SSD tier, so
$\text{T}_\text{i}^{\mathrm{load}}(\text{t})=0$; an SSD-enabled configuration
requires a separately measured loading curve. Requests whose required KV
blocks exceed current HBM capacity remain outside $\mathcal{F}(\text{t})$.

\noindent\textbf{Tiered-Aging parameters.} The evaluated configuration uses
$\tau=5{,}000$~ms and $\text{W}_{\max}=20{,}000$~ms.
After a request waits for $\tau$, an otherwise
unchanged state computational cost is divided by two; $\text{W}_{\max}$ sets the threshold
for oldest-request priority, subject to HBM feasibility. Equal age-adjusted
computational costs are resolved by raw state computational cost, additional KV blocks, and ready-time order. These parameters
are configuration-level controls, not values learned from request arrivals or
tool-call durations, and remain fixed across all evaluated traces.
For another configuration, $\text{W}_{\max}$ can be configured from the
admission-waiting tolerance, while $\tau$ can be calibrated through an
offline sweep on representative workloads to balance \revadd{SLO-compliant throughput} and
admission waiting.

\noindent\textbf{Computational cost calibration for execution.} We profile pure-decode attention
using exact vectors of materialized context lengths on the target
model--hardware--serving-engine configuration. Each measurement compares the
complete-population path with candidate two-cohort paths while excluding the
shared dense/MoE traversal and common KV update. We use two warmup executions,
five measured repetitions, and two passes per shape. Whole context shapes,
rather than individual repetitions, are partitioned across fitting,
calibration, and validation sets. The duration model uses batch size, log total
context length, maximum context length, and context-length coefficient of
variation. The paired gain model additionally uses the context-length ratio,
cut fraction, and per-cohort total and maximum context lengths.

\noindent\textbf{Conservative gain estimate.} When a paired profile is
available, it directly estimates the relative gain of a full-versus-cohort
execution pair and applies a held-out residual bound. Otherwise, \sys derives
the conservative gain from the lower duration bound of
$\widehat{\text{C}}^{\mathrm{full}}(\text{t})$ and the sum of the upper
duration bounds of the two cohorts plus the profiled launch allowance in
$\widehat{\text{C}}_\text{k}^{\mathrm{cohort}}(\text{t})$. Thus, uncertainty
makes cohort execution harder, rather than easier, to select. Missing,
unsupported, or out-of-domain profiles retain complete-population execution.

\noindent\textbf{Cohort parameters.} The implemented controller considers cohort
construction only with at least four participating requests and a
complete-population maximum-to-minimum context-length ratio of at least 1.5.
It uses $\text{M}_{\min}=2$, $\rho=0.35$, an additional-launch allowance of
$0.3$~ms, and $\Gamma=0.05$. Consequently, each cohort contains at least two
requests and at least 35\% of the participating population, and cohort execution
must provide at least 5\% conservative predicted gain after accounting for its
additional launch. Gain ties favor the more balanced split and then the
smaller cut index. These gates are validated with the configuration profile and
remain fixed across workloads.

\noindent\textbf{Workload independence.} Profiling is performed offline, and
\sys performs no online fitting or per-trace retuning. Runtime decisions use
only current \revadd{KV-state residency}, missing-prefix work, ready-wait age, and materialized
context lengths; they do not use future output lengths or future tool-return
times. Workload dynamics therefore change the inputs to the admission decisions
but not the calibrated computational cost functions or their parameters.

\end{document}